%% file: avellarReviewedCSQCDIII.tex
\documentclass[12pt]{article}
\usepackage{epsfig}

\input econfmacros-csqcd3.tex
\def\Title#1{\begin{center} {\Large {\bf #1} } \end{center}}

\begin{document}

\Title{Entropy, Disequilibrium and Complexity in Compact Stars: \\ An information theory approach to understand their Composition}

\bigskip\bigskip


\begin{raggedright}

{\it Marcio G B de Avellar $^{1}$ \\ Jorge E Horvath $^{2}$\index{de Avellar, M.}\\
Instituto de Astronomia, Geof\'isica e Ci\^encias Atmosf\'ericas $^{1,2}$\\
Universidade de S\~ao Paulo\\
05570-010 Cidade Universit\'aria\\
S\~ao Paulo, SP\\
Brazil\\
{\tt Email: marcavel@astro.iag.usp.br $^{1}$ \\ \hspace{1.75cm}foton@astro.iag.usp.br $^{2}$}}
\bigskip\bigskip
\end{raggedright}

\section{Introduction}

The composition of neutron stars is an issue that has been studied for decades. Yet we do not know
exactly what these very compact objects are made of. At this stage of the technological development
the best we can do is to constrain the range of equations of state via the mass-radius diagram.
From the theoretical point of view the things are not easier. The theory of matter at high density and 
temperature is not well established.

Keeping this in mind we applied the Information Theory as a novel way to restrict the range of possible
equations of state for neutron stars. Information Theory is due to Claude E. Shannon in a seminal book in 1949 \cite{Shannon1949}. Basically it deals with the amount of information carried by a string of symbols, or signal, each with its own probability of occurrence. Later, it was shown that Information Theory is a much broader theory from which Statistical Mechanics can be derived.

Information Theory has been successfully applied in other fields of knowledge, like linguistics, genomic, and
condensed matter, to quote only a few. However, the first to apply concepts related to Information Theory
on compact astrophysical objects were Sa\~nudo and Pacheco \cite{Sanudo2009} for white dwarfs, followed by Chatzisavvas {\it et al.} \cite{ChatzisavvasETal2009} for a simple model of neutron stars.

Quantifying the amount of information stored in a neutron star (via the definition of 
{\it information entropy}) we used a statistical measure of {\it complexity} in order to try to classify and 
to infer a hierarchy of equations of state to be realized in Nature.

\begin{figure}[htb]
\begin{center}
\epsfig{file=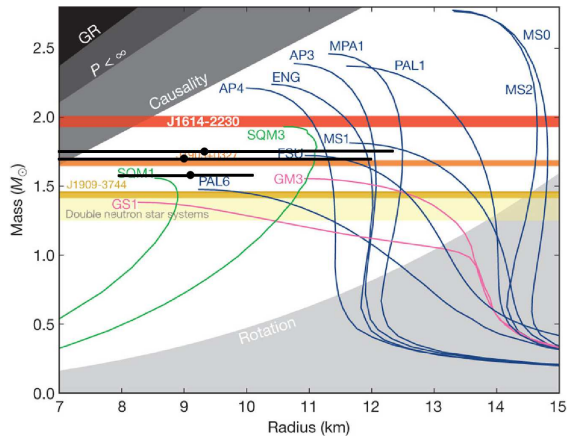,height=2.25in}
\caption{The mass-radius diagram for different EoSs and some constraints. The red strip shows the Demorest pulsar J1614--2230 \cite{Demorest2010}, which has a very precise mass measurement. Adapted from \cite{Demorest2010}.}
\label{fig:massRadiusDem}
\end{center}
\end{figure}

Each equation of state (EoS), e.g. each composition, encodes a univocal sequence of stars in the mass-radius diagram. 
This is our main motivation to think that each EoS encodes also an amount of information in an univocal pattern too. See
Fig.~\ref{fig:massRadiusDem} for a set of sequences of stars with different EoSs and some constraints.

So, the starting point here is how to calculate the information stored in each neutron star of some chosen EoS. We need something that can be the proper analogous to the ``sequence of symbols'' of the Shannon's Information Theory. We can get a hint of what this could be from solving for the structure of a neutron star.

Solving for the structure of a neutron star means that we need to solve the coupled equations of the relativistic hydrostatic equilibrium and the mass continuity for a given EoS. The solution then comes from the Tolman-Oppenheimer-Volkoff equations and provides us with the radial pressure profile, the radial density profile, the radial mass profile, and the radial profiles for each of the elements of the metric inside the star. These profiles can be thought as a kind of distribution along the interior of the stars, analogously (or maybe related) to a continuous probability distribution. 

An appropriate choice appears to be the mass density profile (see next section). Then, because each EoS provides an unique sequence of stars in the mass-radius diagram and because each star of this sequence has its own (and different) density profile, we can use these density profiles to calculate the amount of information stored in each of these stars and compare the information content among stars of the same nature and with different compositions.

The next step is to quantify in some manner something that we can call the statistical complexity of these stars from the information content of each one. We have some intuition of what complexity could mean for a physical system and we will base ourselves on this intuition to think about how Nature would favour some system in relation to other.

Within this approach we expect that, by comparing how the information and complexity change from star to star and from sequence to sequence, we can get some way to ascertain which EoS is preferable in Nature.

Based on these statistical measures, our results point towards to the quarkonic matter as the preferable one.

\section{Calculations}

Let us now define the quantities we want to calculate. The first of them is the {\it complexity}. By complexity we mean what does not match the requirements of being simple. It can be seem tautological, but in physics we always begin with ideal systems as the simplest systems possible and this definition follows our common sense. The other quantity is {\it information}. Information is what we can get from observing the occurrence of an event (how surprising, or unexpected or what else) although it is subject to personal beliefs. To avoid as much as we can arbitrary definitions of information, with a certain reductionism we will define it in terms of the probability of an event to occur.

From some desired mathematical properties of information we can derive:
\begin{equation}
I(p)=-log_{b}(p)
\end{equation} for some probability $p$ and basis $b$ (that gives the unit). $b=2$ give us {\it bits}. For example, 
flipping a fair coin once give you $-log_{2}(1/2)=1$ bit of information.

If a source provides $n$ symbols $\{a_{i}\}$ with probability $\{p_{i}\}$,  then the average amount of information 
in the stream of symbols is:
\begin{equation}
\frac{I}{N}=-\sum_{i=0}^{N}p_{i}log_{b}(p_{i})\equiv H(P).
\end{equation} This quantity is defined as the {\it entropy} of the probability distribution $P=\{p_{i}\}$ and the maximum 
of this quantity is achieved at equiprobability.

Following the reasoning of L\'opez-Ruiz, Mancini and Calbet \cite{LopezRuiz1995}, let us allow complexity to encode order and disorder (or the self-organization of a system) and use two ideal systems, extremes in all aspects and opposites as well: the {\it perfect crystal} and the {\it ideal gas}. The former has zero complexity by definition and has strict symmetry rules in a way that the probability density is centered around one prevailing state of perfect symmetry. Thus, the perfect crystal has minimal information and is completely ordered. The ideal gas also has zero complexity by definition. It has equiprobable accessible states which means maximal information. It is totally disordered.

\begin{figure}[htb]
\begin{center}
\epsfig{file=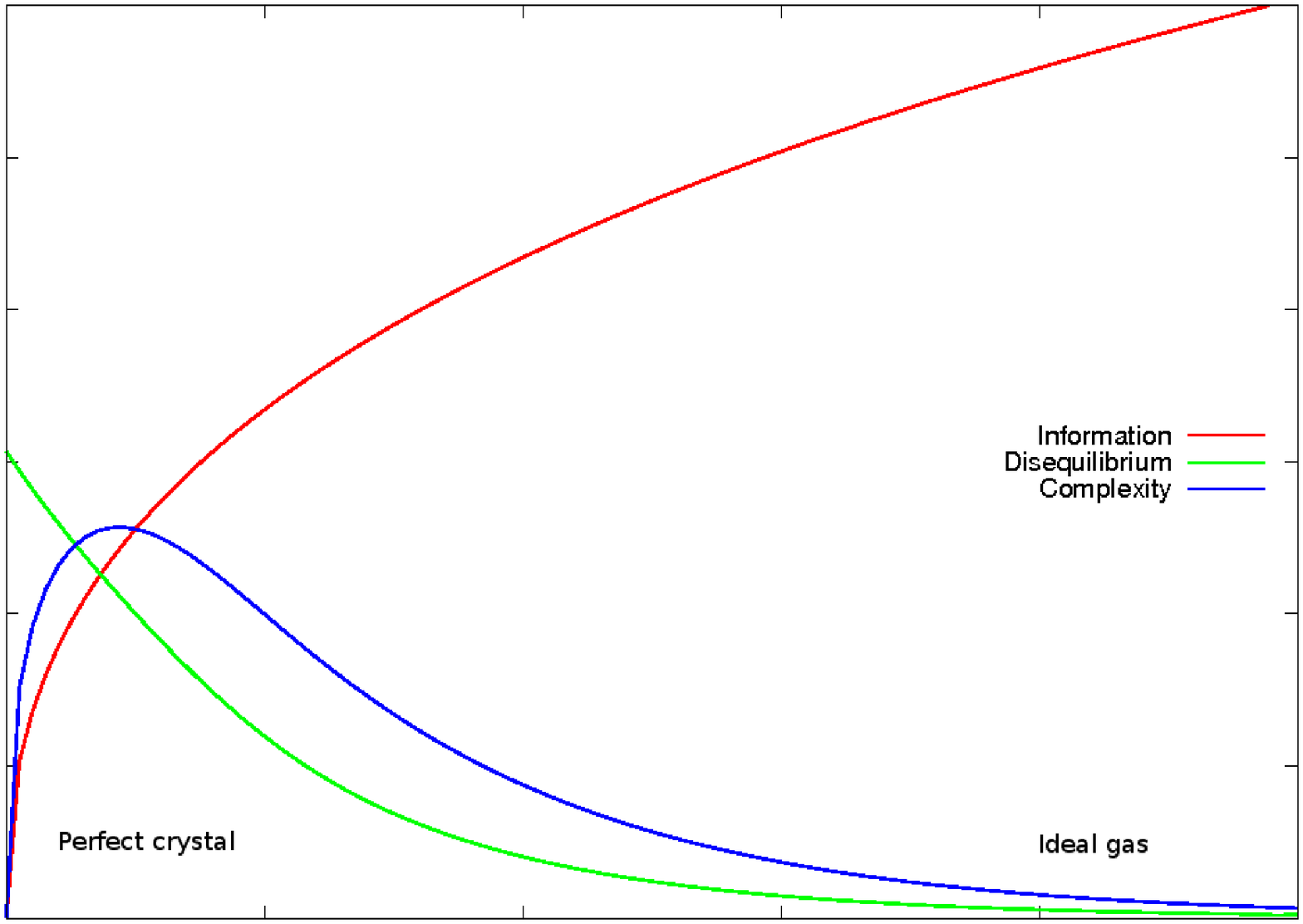,height=2.25in}
\caption{Getting some intuition about complexity (blue curve) with the aid of the information content
of a system (red curve). The disequilibrium (green curve) is the missing ingredient. Adapted from \cite{LopezRuiz1995}.}
\label{fig:intuition}
\end{center}
\end{figure}

We readily see by comparing these two systems that the information alone is not enough to assess what we intend to be
the complexity of a system. Then, before moving on, let us get some intuition of what we expect from something called
complexity. See Fig.~\ref{fig:intuition}. 

From Fig.~\ref{fig:intuition} we see that to get the asymptotic behaviour of complexity (blue curve) for the two ideal systems we
considered previously we need something else than the information entropy (red curve): we define the {\it disequilibrium} (green curve)
as the distance from the state of equiprobability.

The expressions we used are as follow (extended to the continuous case):

\begin{equation}
S=-\int p(x)log_{b}[p(x)]dx
\end{equation}
\begin{equation}
D=\int p^{2}(x)dx
\end{equation}
\begin{equation}
C\equiv S\times D\hspace{0.5cm} or \hspace{0.5cm} C\equiv e^{S}\times D.
\end{equation} where $S$ is the information entropy, $D$ is the disequilibrium and $C$ is the complexity and we 
adopted the last version of complexity for convenience (see Catal\'an, Garay and L\'opez-Ruiz \cite{Catalan2002}).

Having defined all the quantities we wanted we applied to neutron stars with two different compositions, representative of two very different states of matter. We chose a typical hadronic neutron rich composition with SLy4 equation of state (Douchin and Haensel \cite{Haensel2001}) and an equation of state representative of a quark composition with three flavours in equal amounts or {strange quark matter} (Witten \cite{Witten1984}).

How does the composition affect the measures of these quantities? But first, what should we adopt as $p(x)$? 

As $p(x)$ we used the energy density distribution, $\epsilon(r)$, obtained by calculating the relativistic internal structure of
the neutron stars given by the Tolman-Oppenheimer-Volkoff equations. This is plausible because the energy density is related 
to the probability of finding some particles at a certain location inside the star, so it is a suitable measure of ``probability''.

For each star of our two sequences, e.g. given by each calculated density profile, we performed the integrals, multiplying the result
by an appropriate constant that makes the quantities dimensionless.


\section{Results}

\begin{figure}[htb!]
\begin{center}
\epsfig{file=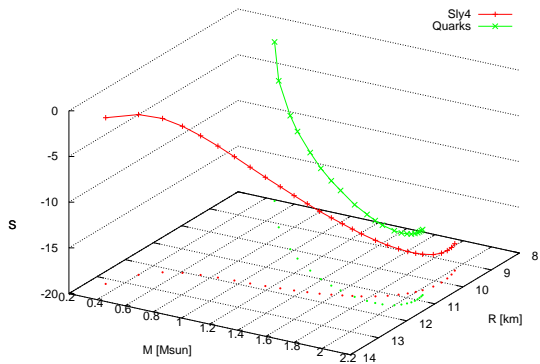,height=2.25in}
\caption{Information entropy ($S$) $\times$ mass $\times$ radius. Higher the masses, smaller entropies.}
\label{fig:entropy}
\end{center}
\end{figure}

\begin{figure}[htb!]
\begin{center}
\epsfig{file=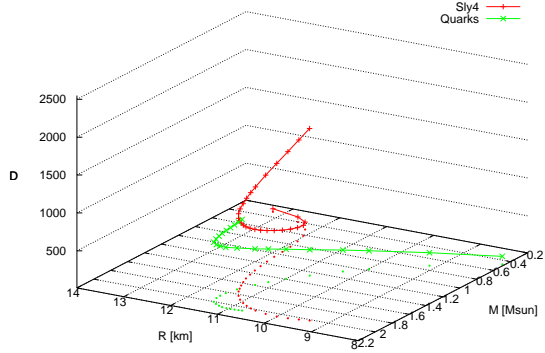,height=2.25in}
\caption{Disequilibrium ($D$) $\times$ mass $\times$ radius. Higher the masses, higher disequilibria.}
\label{fig:disequilibrium}
\end{center}
\end{figure}

\begin{figure}[htb!]
\begin{center}
\epsfig{file=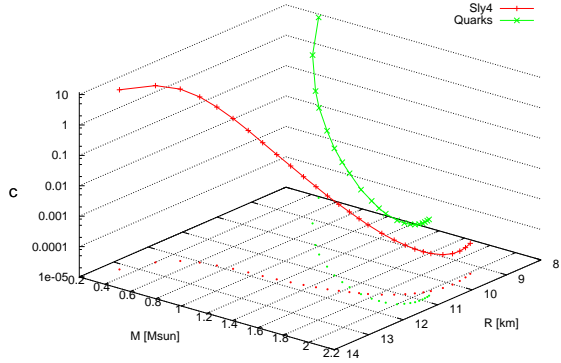,height=2.25in}
\caption{Complexity ($C$) $\times$ mass $\times$ radius. Higher the masses, smaller complexities.}
\label{fig:complexity}
\end{center}
\end{figure}

In Figs.~\ref{fig:entropy},~\ref{fig:disequilibrium} and~\ref{fig:complexity} we show the results of the measures of the information entropy, 
disequilibrium and complexity as a function of the mass and radius of each star of the sequences.

Our results (de Avellar and Horvath \cite{deAvellar2012}) are in rough agreement with Chatzisavvas {\it et al.} \cite{ChatzisavvasETal2009}.

\section{Conclusions}

Our the results show that hadronic stars have low complexity and are ordered systems, tending to the perfect crystal. On the other hand, quark stars also have low complexity but are less ordered systems because they have lower disequilibrium being more distant from perfect crystal.

Sa\~nudo and Pacheco \cite{Sanudo2009} performed these calculations for the white dwarf case: they found that the complexity {\it grows} with increasing mass, reaching a maximum finite value at the Chandrasekhar mass. Their results show a remarkable resemblance with atomic systems if one
substitute the mass by the atomic number.

From our results, we can state three main conclusions. First, if order costs energy, then nature should favour exotic strange quark stars. Also, there is a trend for these stars to be at a state of minimum complexity. Calbet and L\'opez-Ruiz \cite{Calbet2007}, \cite{Calbet2001} have shown that for a system out of equilibrium there is, in fact, a tendency of the complexity to reach an extremum. If a transition hadronic $\rightarrow$ quark occur, that would be the case. At last, the entropy trend as a function of compactness also points towards the self-bound quark stars, although there is an entropy barrier that has to be overcome which is an issue that our group has a study in progress.

\end{document}

%% file: econfmacros-csqcd3.tex
\def\beq{\begin{equation}}
\def\eeq#1{\label{#1}\end{equation}}
\def\eeqn{\end{equation}}

\def\beqa{\begin{eqnarray}}
\def\eeqa#1{\label{#1}\end{eqnarray}}
\def\eeqan{\end{eqnarray}}

\let\bar=\overbar

\def\Dslash{\not{\hbox{\kern-4pt $D$}}}
\def\dslash{\not{\hbox{\kern-2pt $\del$}}}

\def\msb{{\bar{\ssstyle M \kern -1pt S}}}

\usepackage{fancyhdr,graphicx}